\pdfoutput=1
\documentclass[twocolumn,prl]{revtex4} 
\usepackage{graphicx} 
\usepackage{bbm} 
\usepackage{amsmath} 
\usepackage{amssymb}
\usepackage{color}

\def\braket#1{\mathinner{\langle{#1}\rangle}}
\def\bra#1{\left\langle#1\right|}
\def\ket#1{\left|#1\right\rangle}

\newcommand{\be}{\begin{equation}} 
\newcommand{\ee}{\end{equation}} 
\newcommand{\bea}{\begin{eqnarray}} 
\newcommand{\eea}{\end{eqnarray}}

\renewcommand{\epsilon}{\varepsilon}  

\begin{document} 
 
\title{Multi-level interference resonances in strongly-driven three-level systems}

\author{J. Danon and M. S. Rudner}
\affiliation{ 
Niels Bohr International Academy, and the Center for Quantum Devices, Niels Bohr Institute, University of Copenhagen, 2100 Copenhagen, Denmark
} 
 
 
\begin{abstract} 
We study multi-photon resonances in a strongly-driven three-level quantum system, where one level is periodically swept through a pair of levels with constant energy separation $E$.
Near the multi-photon resonance condition $n\hbar\omega = E$, where $n$ is an integer, we find qualitatively different behavior for $n$ even or odd.
We explain this phenomenon in terms of families of interfering trajectories of the multi-level system. 
Remarkably, the behavior is insensitive to fluctuations of the energy of the driven level, and survives deep into the strong dephasing regime.
The setup can be relevant for a variety of solid state and atomic or molecular systems.
In particular, it provides a clear mechanism to explain recent puzzling experimental observations in strongly-driven double quantum dots.
\end{abstract} 
 
\maketitle 

The advent of intense microwave and laser sources has opened a range of new possibilities for investigating the strong-driving regime of both natural and artificial (solid-state) atoms and molecules \cite{PhysRevA.76.042514,lang:natphys,Oliver:science}.
In this regime, the amplitude of an applied ac driving field may greatly exceed both the driving field photon energy, $\hbar\omega$, as well as the separation between energy levels of the system.
High-order multi-photon processes and multi-level coherences may then become important \cite{scullyzubairy,sun:natcomm}, leading to interesting dynamical effects which go well beyond the canonical Rabi oscillations of weakly driven two-level systems.

While the dynamics of strongly-driven two-level systems have been studied extensively, both theoretically \cite{PhysRevA.75.063414,lzs_review,PhysRevB.87.235318} and experimentally \cite{Oliver:science,PhysRevB.86.121303}, {\it multi}-level systems offer new avenues to explore.
Intriguing and potentially useful phenomena such as amplitude spectroscopy \cite{berns:nature}, population inversion \cite{Sun:APL, deGraaf:PRL, reilly:natcomm}, and microwave-induced cooling \cite{Valenzuela08122006} have been realized in a variety of systems.

Recently a new type of multi-photon resonance was discovered in experiments on spin-blockaded double quantum dots (DQDs)
subjected to large-amplitude modulations of a nearby gate electrode \cite{Laird2009,PhysRevLett.112.227601}.
The resonances show a striking asymmetry, with current {\it enhanced} when the electron Zeeman splitting matches an odd-integer multiple of the driving field photon energy, $E_Z = (2n + 1)\hbar\omega$, and {\it suppressed} for even-integer resonances $E_Z = 2n\hbar\omega$.
Such a dramatic even/odd effect does not occur in two-level systems, and appears to be a robust feature of the multi-level DQDs.
Analytical~\cite{PhysRevB.84.241305,PhysRevB.89.115409} and numerical~\cite{PhysRevB.86.125428,stehlik:v1} investigations have accounted for the existence of multi-photon resonances, but crucially could not explain the even/odd asymmetry (though a Fano-like origin was speculated~\cite{PhysRevB.84.241305}).

Motivated by this puzzle, we look at the dynamics of strongly-driven multi-level systems.
For the case of three levels we find multi-photon resonances with characteristics which differ markedly from those of familiar two-level resonances.
We connect our model to the experiments of \cite{Laird2009,PhysRevLett.112.227601} and show that it captures all relevant features of the data, including the striking even/odd asymmetry.
\begin{figure}[b]
\begin{center}
\includegraphics[width=82mm]{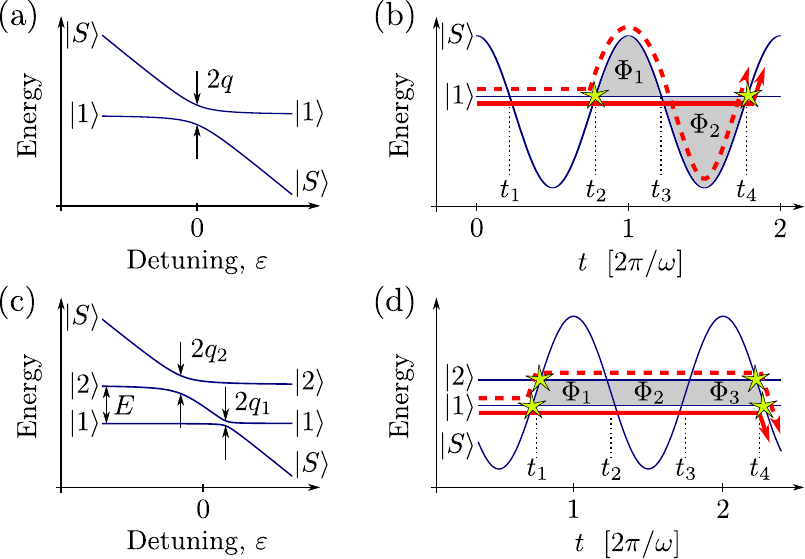}
\caption{(Color online) Spectrum of (a,b) the two-level Hamiltonian (\ref{eq:2lh}) and (c,d) the three-level Hamiltonian (\ref{eq:3lh}). In (a,c) the energy levels are plotted as a function of detuning and in (b,d) as a function of time assuming strong driving (thin blue lines). We have set: (b) $A = 5 \varepsilon \gg q$, and (d) $A = 3E \gg q_{1,2}$, $\varepsilon = 0$. In (b,d) we added pairs of possible paths in time bringing the system from $\ket{1}$ to $\ket{S}$ (thick red lines). The paths shown in (d) illustrate why we expect resonances at $E=2n\omega$.}\label{fig:fig1}
\end{center}
\end{figure}

To highlight the key qualitative differences between two-level and multi-level resonances, we first briefly review the phenomenology of multi-photon resonances in a two-level system.
We consider a system with basis states $\{ \ket{1}, \ket{S} \}$, its dynamics governed by the Hamiltonian
\begin{equation}
H_2(t) = \left( \begin{array}{cc} 0 & q \\ q & -\epsilon(t) \end{array} \right), \quad \epsilon(t) = \varepsilon_0 - A \cos \omega t.
\label{eq:2lh}
\end{equation}
Here we focus on the case of strong driving, $A \gg q$ and $A > |\epsilon_0|$.
Figures \ref{fig:fig1}a,b show the instantaneous spectrum of this system, plotted versus detuning $\epsilon$, and time $t$. 

The relevant features of the driven system's dynamics can be understood heuristically in terms of families of interfering trajectories (Fig.~\ref{fig:fig1}b).
For strong driving, transitions take place at relatively well defined points in time $\{t_p\}$ when the two levels are nearly degenerate.
Two paths taking the system from state $\ket{1}$ to $\ket{S}$ are indicated by the dashed and solid red lines.
In the illustration, the transitions occur at times $t_2$ and $t_4 = t_2 + T$, where $T = 2\pi/\omega$ is the driving period, and the interference phase corresponds to the difference of shaded areas shown, $\Phi = |\Phi_1| - |\Phi_2|$. 
When $\epsilon_0 = n\omega$ (we set $\hbar = 1$), we have $\Phi = 2n\pi$, and for integer $n$ the interference is constructive.
In this case all paths featuring transitions at ``even'' times $t_{2p}$ mutually interfere constructively, as do all paths with transitions at odd times $t_{2p+1}$.
This provides a resonant response.
Additional structure results from interferences between these two groups of trajectories, which are sensitive to the individual phases $\Phi_{1,2}$.
In the case of sinusoidal driving, the resulting $\Phi_{1,2}$ give rise to the characteristic ``Bessel staircase'' of modulated resonance intensities, with the Bessel function $J_n(A/\omega)$ controlling the strength of the $n$-photon resonance line \cite{lzs_review}.
The intensities of these two-level multi-photon resonances are thus highly sensitive to both the amplitude and frequency of driving, exhibiting sequences of peaks and nodes as $A/\omega$ is varied.

We now turn our attention to strong driving in a {\it multi-level} system.
To clearly demonstrate the essential physics of multi-level resonances, we focus on the case of three levels.
We assume that the driving field couples strongly to one level, $\ket{S}$, while the energy separation between the other levels $\ket{1}$ and $\ket{2}$ is unaffected (see Figs.~\ref{fig:fig1}c,d).
The state $\ket{S}$ therefore acts as a ``shuttle,'' mediating population transfer between $\ket{1}$ and $\ket{2}$.
This situation is described by the generic Hamiltonian
\begin{align}
H_3(t) = \left( \begin{array}{ccc} E/2 & 0 & q_2 \\ 0 & -E/2 & q_1 \\ q_2 & q_1 & -\epsilon(t) \end{array} \right),
\label{eq:3lh}
\end{align}
written in the basis $\{\ket{2}, \ket{1}, \ket{S}\}$. 
Here, $E$ is the energy splitting between states $\ket{1}$ and $\ket{2}$, 
$q_{1,2}$ are the coupling matrix elements, and $\epsilon(t)= \varepsilon_0 - A \cos \omega t$ as before.

Two-level resonances between $\ket{S}$ and $\ket{1}$ or $\ket{2}$, analogous to those described above, can occur whenever the corresponding static detuning $\epsilon_0 \pm E/2$ matches the $n$-photon energy $n\omega$.
Such resonances do not present qualitatively new physics.

More interestingly, we investigate the existence of resonances associated with the energy splitting $E$.
Such resonances must occur via the strongly modulated level $\ket{S}$, thereby constituting a true multi-level phenomenon.

How could such resonances arise?
In the spirit of the discussion above, in Fig.~\ref{fig:fig1}d we illustrate a characteristic pair of interfering trajectories, in this case from $\ket{1}$ to $\ket{S}$.
For large driving amplitude $A \gg \epsilon_0, E$, the interference phase is given by $\Phi_1 + \Phi_2 + \Phi_3 \approx E\,(t_4 - t_1) = 3 E \, (T/2)$.
Importantly, this phase is controlled only by the splitting $E$ and the driving half-period $T/2=\pi/\omega$, and {\it not} by the driving amplitude or waveform.
There exist many such paths, where the last two transitions take place approximately at the same time $t_{p>1}$, all contributing to the full transition amplitude at the same (fourth) order in the couplings $q_{1,2}$.
Constructive interference for this series of paths is achieved when $\Phi_1 = \pi E / \omega = 2\pi n$, suggesting the existence of resonances at $E = 2n\omega$, i.e.~at {\it even} multiples of $\omega$.
Similar considerations for transitions from $\ket{1}$ to $\ket{2}$ reveal a series of processes depending on the {\it full} driving period $T$, predicting additional resonances at {\it all} multiples of the photon energy, $E = n\omega$.
Thus we expect this system to display resonances for driving frequencies commensurate with the splitting $E$, showing very different behavior for $E$ an even or odd multiple of $\omega$.
Further, in sharp distinction with the two-level case discussed above, these resonances are only weakly sensitive to the driving amplitude $A$ and detuning $\varepsilon_0$. Indeed, the interference phase $\Phi_1$ only changes appreciably when $\varepsilon_0$ is varied on the order of $A$, or vice versa.

We now begin our detailed analysis, which is based on a perturbative treatment in terms of the small parameters $q_{1,2}^2/(A\omega)$ that characterize the strong driving limit.
To most clearly exhibit the effect, and to allow us to arrive at analytic results, we focus on a regime of strong dephasing where coherences between $\ket{1}$ and $\ket{S}$ and between $\ket{2}$ and $\ket{S}$ are rapidly lost, on a time scale shorter than the driving period.
In contrast, we allow coherences between $\ket{1}$ and $\ket{2}$ to be long-lived on this timescale.
The dephasing is modeled by Gaussian white-noise fluctuations on each of the unperturbed energy levels via
\begin{align}
\delta H_3 (t)= \sum_{\alpha} \xi_\alpha(t) \ket{\alpha}\bra{\alpha},\quad \alpha \in \{1, 2, S\},
\end{align}
with $\overline{\xi_\alpha (t) \xi_\beta (t')} = \Gamma_\alpha \delta(t-t')\delta_{\alpha\beta}$, where the overbar indicates averaging over noise realizations.
Within this model we calculate the rates of interlevel transitions, working up to fourth order in the couplings $q_{1,2}$.

Strong dephasing is particularly relevant for the experiments in Refs.~\cite{Laird2009,PhysRevLett.112.227601}, where the level corresponding to $\ket{S}$ exhibits strong lifetime broadening due to coupling to a nearby reservoir (see discussion below).
The multi-level resonances survive deep into the strong-dephasing regime, where the quasi-two-level resonances at $\epsilon_0 \pm E/2 = n\omega$ are completely washed out.

The first analytical step is to transform to a modified interaction picture via $\ket{\psi_R(t)} = e^{iR(t)}\ket{\psi(t)}$, with $R(t) = \sum_{\alpha} \phi_\alpha(t)\ket{\alpha}\bra{\alpha}$.
The phases $\phi_\alpha$ are given by  $\phi_\alpha(t) = -\int_0^t d\tau\, \tilde{\epsilon}_\alpha(\tau)$, with $\tilde\epsilon_{1,2}(\tau) = \mp\frac12 E + \xi_{1,2}(\tau)$ and $\tilde\epsilon_S(\tau) = \epsilon(\tau) + \xi_S(\tau)$.
States in this interaction picture evolve according to $i\frac{d}{dt}\ket{\psi_R} = \tilde{H}_3(t)\ket{\psi_R}$, with ${\tilde{H}_3(t) = -\dot{R} + e^{iR(t)}(H_3+\delta H_3)e^{-iR(t)}}$.
This yields
\begin{align}
\label{eq:H3R}\tilde{H}_{3}(t) = \left( \begin{array}{ccc} 0 & 0 & q_2 e^{-i\phi_{S2}(t)} \\ 0 & 0 & q_1 e^{-i\phi_{S1}(t)} \\ q_2 e^{i\phi_{S2}(t)} & q_1 e^{i\phi_{S1}(t)} & 0 \end{array} \right),
\end{align}
where $\phi_{\alpha\beta}(t) \equiv \phi_\alpha(t) - \phi_\beta(t)$.

The transition rate between states $\ket{\alpha}$ and $\ket{\beta}$ is calculated as the time-derivative of the transition probability,
\begin{align}
W_{\alpha\to\beta} = \frac{d}{dt} \overline{\big|{\braket{\beta|U(t)|\alpha}}\big|^2},
\end{align}
where $U(t)$ evolves the system between times $0$ and $t$.
We expand the time-evolution operator in powers of $q_{1,2}$ as $U(t) = 1 + U^{(1)}(t) + U^{(2)}(t) + \dots$, with
\begin{align}
U^{(m)}(t) & = (-i)^m\int_0^t dt_1\cdots\int_0^{t_{m-1}}\!\!\!\!\!\!\!dt_m\ \tilde{H}_{3}(t_1)\cdots\tilde{H}_{3}(t_m). \nonumber
\end{align}
Working up to third order in $q_{1,2}$ gives access to the transition {\it rates} up to fourth order in the couplings.

For illustration we now evaluate $W_{1\to S}$ to lowest (second) order; other rates are obtained similarly. 
We write
\begin{align}
W_{1\to S}^{(2)} & = \frac{d}{dt} \overline{\big|{\braket{S|U^{(1)}(t)|1}}\big|^2} \nonumber\\
& = q_1^2 \frac{d}{dt} \int_0^t dt_1\int_0^t dt_2\ \overline{e^{i[\phi_{S1}(t_1)-\phi_{S1}(t_2)]}},
\end{align}
and use $\overline{\exp\{i\xi(t)\}} = \exp\{-\tfrac{1}{2}\overline{\xi(t)^2}\}$.
The result is simplified under the assumption $\Gamma_S \gg \omega,\Gamma_{1,2}$, giving
\begin{align}
W_{1\to S}^{(2)} & = \frac{q_1^2\Gamma_S}{(\frac12 E - \varepsilon_0 + A \cos \omega t)^2 + \tfrac{1}{4}\Gamma_S^2}.
\label{eq:w212}
\end{align}
Moving to the strong driving limit $A \gg \Gamma_S$ and assuming $A>|\epsilon_0 - \tfrac{1}{2}E|$, the transition rate displays sharp bursts, well-separated in time, occurring whenever the levels cross, i.e.~when $A\cos\omega t \approx \varepsilon_0 - \frac12 E$.
Averaging these bursts over one period yields
\begin{align}
W_{1\to S}^{(2)} & \approx \frac{2q_1^2}{\sqrt{A^2 - (\frac12 E - \varepsilon_0)^2}}.
\label{eq:w122nd}
\end{align}
Similarly, we find $W_{2\to S}^{(2)} \approx 2q_2^2/[A^2 - (\tfrac{1}{2}E +  \varepsilon_0)^2]^{1/2}$ in the same limit, and identical rates for the reverse processes $W^{(2)}_{S\to 1}$ and $W^{(2)}_{S\to 2}$.

Multi-level interference resonances first arise at fourth order,
\begin{align}
W^{(4)}_{\alpha\to\beta} = \frac{d}{dt}& \bigg\{ \overline{ \big|{\braket{\beta|U^{(2)}(t)|\alpha}}\big|^2} \nonumber\\
& +\ 2\,{\rm Re}\overline{ \braket{\alpha|U^{\dagger(3)}(t)|\beta}\braket{\beta|U^{(1)}(t)|\alpha}} \bigg\}.\label{eq:w4}
\end{align}
Due to the form of $\tilde{H}_3$, the rates $W^{(4)}_{1\leftrightarrow 2}$ only involve the first term in Eq.~(\ref{eq:w4}), while the rates $W^{(4)}_{1,2 \leftrightarrow S}$ involve only the last. 
Proceeding along similar lines as above, we assume $\Gamma_S \gg \omega,\Gamma_{1,2}$ and work in the strong-driving limit $A\gg \Gamma_S$.

After some algebra we find analytic approximations for the rates in two important cases, valid for times $t \gg \Gamma_{1,2}^{-1}$.
First, at zero detuning, $\varepsilon_0 = 0$, we find $W^{(4)}_{1,2 \leftrightarrow S} \approx - g_0 \bar{W}$ and $W^{(4)}_{1\leftrightarrow 2} \approx (\tfrac{1}{2} g_0 + h_0)\bar{W}$, with $\bar{W} = 2\pi q_1^2q_2^2/(A^2\omega)$ and
\begin{align}
g_0 & = \frac{2\cos(n\pi)\sinh(\tfrac{1}{2}\Gamma')+e^{\Gamma'}-\cos 2n\pi}{\cosh\Gamma'-\cos 2n\pi}, \nonumber\\
h_0 & = \frac{\sin^2(\tfrac{1}{2}n\pi )\sinh(\Gamma')}{\cosh\Gamma'-\cos 2n\pi}.\nonumber
\end{align}
Here we use the (continuum-valued) dimensionless energy splitting $n = E/\omega$ and dephasing rate $\Gamma' = (\Gamma_1+\Gamma_2)\pi/\omega$. 
Second, for arbitrary detuning but {\it integer} $n$, we find $W^{(4)}_{1,2 \leftrightarrow S} \approx -g_i \bar{W}$ and $W^{(4)}_{1\leftrightarrow 2} \approx (\tfrac{1}{2} g_i + h_i)\bar{W}$, with
\begin{align}
g_i & = \frac{\cos(n d_-)[\sinh (\tfrac{\Gamma'd_-}{2\pi}) + \sinh (\frac{\Gamma'd_+}{2\pi})]+e^{\Gamma'}-1}{(\cosh\Gamma'-1)(1-\delta^2)}, \nonumber\\
h_i & = \frac{\sin^2(\tfrac{1}{2}nd_+)\coth(\tfrac{1}{2}\Gamma')}{1-\delta^2},\nonumber
\end{align}
where $d_\pm = \pi \pm 2 \sin^{-1}\delta$ and $\delta = \varepsilon_0/A$.

\begin{figure}
\begin{center}
\includegraphics[width=82mm]{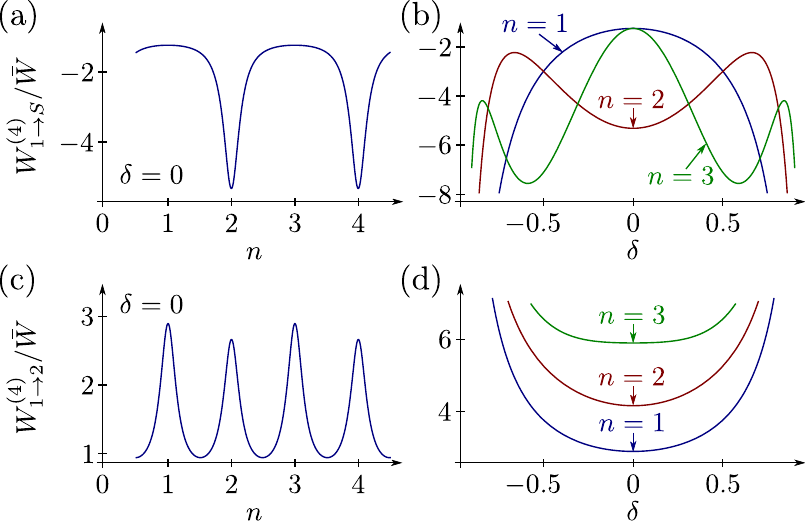}
\caption{(Color online) The rate $W^{(4)}_{1\to S}$ at (a) zero detuning as a function of $n = E/\omega$, and (b) for integer $n=1,2,3$ as a function of $\delta$. (c,d) The same for the rate $W^{(4)}_{1\to 2}$. In all plots we used $\Gamma'/\pi = 0.3$. In (d) the curves are offset in steps of~$\tfrac{3}{2}$.
\vspace{-0.2 in}
}\label{fig:fig3}
\end{center}
\end{figure}
In Fig.\ \ref{fig:fig3} we plot the rates as a function of $n$ for $\delta = 0$ (a,c) and as a function of $\delta$ for $n = 1,2,3$ (b,d), in all plots setting $\Gamma'/\pi = 0.3$. 
The rates display resonant features at integer $n$.
Moreover, the resonances for even and odd $n$ are qualitatively different, as anticipated above. 
The negative sign of $W^{(4)}_{1\to S}$ indicates that this fourth-order contribution provides a suppression of the large (second-order) background transition rate $W^{(2)}_{1 \to S}$, see Eq.~(\ref{eq:w122nd}).
As long as $\Gamma_1+\Gamma_2 > W^{(2)}_{1 \to S}$ the total rate  $W^{(2)}_{1 \to S} + W^{(4)}_{1 \to S}$ is positive. 
For $W^{(2)}_{1 \to S} > \Gamma_1+\Gamma_2$, lifetime broadening of $\ket{1}$ and $\ket{2}$ due to driving-induced transitions to $\ket{S}$ becomes dominant. 
To capture this effect, higher terms in the perturbation expansion must be included.

We now connect our results to the experimental observations of Refs.~\cite{Laird2009,PhysRevLett.112.227601}, in which current through spin-blockaded DQDs was measured in the presence of strong ac driving. 
In the two-electron regime, the low-energy electronic subspace of the DQD is spanned by five states: a ``(1,1)'' spin-singlet and a spin triplet with a single electron in each dot, and a ``(0,2)'' spin-singlet with double occupancy of the right dot (the left dot being empty).
Current flow is mediated by the (0,2) singlet state, which is the only state with direct coupling to the drain lead.

In spin blockade, current is limited by the lifetimes of the  (1,1) triplet states, which to zeroth approximation do not couple to the (0,2) singlet.
Finite coupling between spin triplet and singlet levels occurs via spin-orbit, hyperfine, and/or inhomogeneous Zeeman coupling.
Away from singlet-triplet degeneracy points, ac driving (e.g.~applied to one of the gate electrodes controlling the DQD potential) can provide the energy necessary to stimulate triplet-singlet transitions \cite{katja:science,laird}.
When the driving frequency and level splittings are in resonance, such coupling is expected to lift the blockade and produce an {\it enhancement} of current. The striking even/odd effect observed in Refs.~\cite{Laird2009,PhysRevLett.112.227601} thus clearly does not fit in this simple picture. Furthermore, only a smooth modulation of resonance intensity with $A/\omega$ was observed, in stark contrast to the nodal structure expected for conventional multi-photon resonances as described in the introduction.

As we will now show, the multi-level multi-photon resonances described above account for all of the main features of the experimental data.
The three-level model certainly does not provide a complete representation of dynamics in the full five-dimensional low-energy subspace of a spin-blockaded DQD, but it nonetheless captures the essential physics at play near the resonances.
To make the connection explicit, state $\ket{S}$ in our model represents the (0,2) singlet state of the DQD, while $\ket{1}$ represents the triplet state ${T_+}$, with both electron spins pointing up, and $\ket{2}$ represents a particular superposition of the (1,1) singlet and $T_0$ states, which is determined by Zeeman energy inhomogeneities in the DQD \cite{footnote_fourthlevel}.

Using all contributions to the transition rates up to fourth order, we compute the steady-state current via a master equation for the time-dependent level occupation probabilities $\{p_\alpha\}$,
\begin{align}
\frac{dp_1}{dt} = & -p_1(W_{1\to S}+W_{1\to 2}) + p_2W_{2\to 1} \nonumber\\
\label{eq:Master}& + \frac{1}{2}(p_1W_{1\to S}+p_2W_{2\to S}),
\end{align}
where $p_2 = 1-p_1$.
To eliminate $p_S$, we assumed that the decay of $\ket{S}$ and the consecutive reloading of $\ket{1}$ or $\ket{2}$ (with equal probabilities) happens instantaneously on the time scale of the dynamics of $p_{1,2}$. 
We solve for $dp_1^{\rm (eq)}/dt=0$, with the steady-state current following as $I/e = p_1^{\rm (eq)}W_{1\to S} + p_2^{\rm (eq)}W_{2 \to S}$.
\begin{figure}
\begin{center}
\includegraphics[width=82mm]{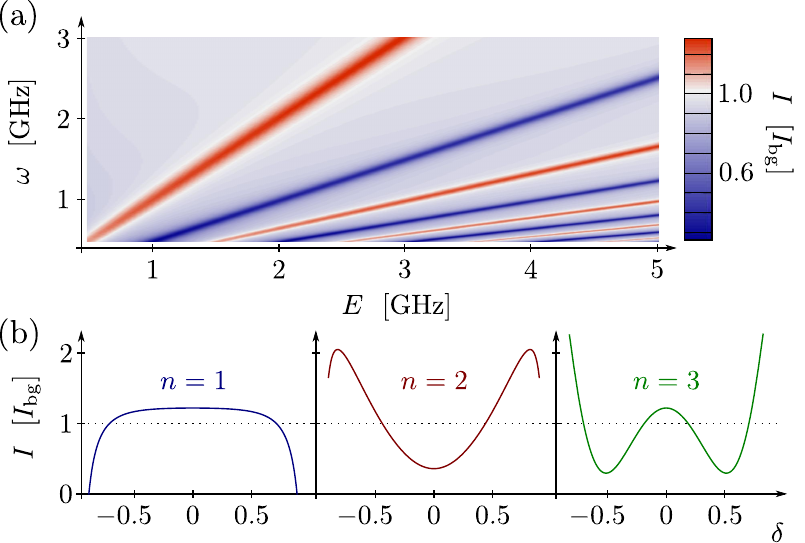}
\caption{(Color online) Calculated current through a driven double quantum dot in spin blockade, normalized to the background current $I_{\rm bg}$.
(a) The current at $\delta = 0$ as a function of $\omega$ and $E$. (b) Slow modulation of the resonances: the current as a function of $\delta$ for fixed $n=1,2,3$. In all plots we used $q_1^2/A = 5$~MHz, $q_2^2/A = 50$~MHz, and $\Gamma_{1,2} = 100$~MHz.
\vspace{-0.2 in}
}\label{fig:fig4}
\end{center}
\end{figure}

To compare with the data presented in Fig.~2d of Ref.~\cite{PhysRevLett.112.227601}, we set $\delta = 0$ and assume that $\omega, E \sim$ GHz.
We set $q_1^2/A = 5$~MHz and $q_2^2/A = 50$~MHz, i.e.~$q_2^2 / q_1^2 = 10$~\cite{PhysRevB.81.201305}, and choose $\Gamma_{1,2} = 100$~MHz.
In Fig.~\ref{fig:fig4}a we plot the resulting steady-state current, normalized to $I_{\rm bg}$, the off-resonant ``background'' current (i.e.~the current due to ``direct'' second-order transitions $W_{1,2 \to S}^{(2)}$ associated with repeated sweeps through the $S$-$T_+$ level crossing; in the experiment $I_{\rm bg} \sim 15$~pA).
The model reproduces all important features of the data: a resonant response of current along all $n$-photon lines, alternating between enhancement for odd $n$ and suppression for even $n$. 
At even $n$, the negative contributions $W_{1,2 \to S}^{(4)}$ suppress escape from $\ket{1}$ and $\ket{2}$ to $\ket{S}$, resulting in a reduction of current relative to the background.
The rate $W_{1 \leftrightarrow 2}^{(4)}$ is largest for odd $n$, where it efficiently mixes $\ket{1}$ and $\ket{2}$ and thus enhances the escape rate out of the most strongly blocked state, $\ket{1}$, thereby increasing the total current.
Including a second unpolarized (1,1) level~\cite{footnote_fourthlevel}, split from $\ket{1}$ by $E'$, would yield another fan of current peaks and dips at $E'=n\omega$, reproducing the ``doubled'' line shape of Fig.~2d of Ref.~\cite{PhysRevLett.112.227601}.

We finally investigate the detuning-dependence of the current, which in the experiment showed a strikingly slow modulation (on the scale of $\varepsilon_0\sim A$) with qualitatively distinct shapes for each of the resonances, see Fig.~3b of \cite{PhysRevLett.112.227601}. 
In Fig.~\ref{fig:fig4}b we plot the current as a function of $\delta$ at fixed $n = 1,2,3$, using the same parameters as for Fig.~\ref{fig:fig4}a. 
The detuning-dependence of $I$ agrees well with the experimental observations.
Here it arises from the weak dependence of the interference phases $\Phi_n$ on $\varepsilon_0$, as explained above.

To summarize, we investigated multi-photon resonances in a strongly-driven three-level quantum system. We identified new resonant responses which crucially depend on the multi-level structure of the system. We further revealed how these resonances provide a mechanism to explain recent puzzling experimental observations in strongly-driven double quantum dots.
Interestingly, the behavior survives deep into the regime of strong dephasing on one of the levels.
Detailed explorations of the fully-coherent regime and the role of decoherence are interesting directions for further study.

We thank C.~M. Marcus and K. Flensberg for helpful discussions.
MR acknowledges support by the Villum Foundation.


\begin{thebibliography}{10}

\bibitem{PhysRevA.76.042514}
M.~Mark, F.~Ferlaino, S.~Knoop, J.~G. Danzl, T.~Kraemer, C.~Chin, H.-C.
  N\"agerl, and R.~Grimm, Phys. Rev. A \textbf{76}, 042514 (2007).

\bibitem{lang:natphys}
F.~Lang, P.~{v. d. Straten}, B.~Brandstätter, G.~Thalhammer, K.~Winkler, P.~S.
  Julienne, R.~Grimm, and J.~{Hecker Denschlag}, Nat. Phys. \textbf{4}, 223
  (2008).

\bibitem{Oliver:science}
W.~D. Oliver, Y.~Yu, J.~C. Lee, K.~K. Berggren, L.~S. Levitov, and T.~P.
  Orlando, Science \textbf{310}, 1653 (2005).

\bibitem{scullyzubairy}
M.~O. Scully and M.~S. Zubairy, \textit{Quantum Optics} (Cambridge University Press, Cambridge, UK, 1997).
  
\bibitem{sun:natcomm}
G.~Sun, X.~Wen, B.~Mao, J.~Chen, Y.~Yu, P.~Wu, and S.~Han, Nat. Comm.
  \textbf{1}, 51 (2010).

\bibitem{PhysRevA.75.063414}
S.~Ashhab, J.~R. Johansson, A.~M. Zagoskin, and F.~Nori, Phys. Rev. A
  \textbf{75}, 063414 (2007).

\bibitem{lzs_review}
S.~N. Shevchenko, S.~Ashhab, and F.~Nori, Phys. Rep. \textbf{492}, 1 (2010).

\bibitem{PhysRevB.87.235318}
H. Ribeiro, J.~R. Petta, and G. Burkard, Phys. Rev. B \textbf{87}, 235318 (2013).

\bibitem{PhysRevB.86.121303}
J.~Stehlik, Y.~Dovzhenko, J.~R. Petta, J.~R. Johansson, F.~Nori, H.~Lu, and
  A.~C. Gossard, Phys. Rev. B \textbf{86}, 121303 (2012).

\bibitem{berns:nature}
D.~M. Berns, M.~S. Rudner, S.~O. Valenzuela, K.~K. Berggren, W.~D. Oliver,
  L.~S. Levitov, and T.~P. Orlando, Nature \textbf{455}, 51 (2008).

\bibitem{Sun:APL}
Guozhu Sun, Xueda Wen, Yiwen Wang, Shanhua Cong, Jian Chen, Lin Kang, Weiwei Xu, Yang Yu, Siyuan Han, and Peiheng Wu,
Appl. Phys. Lett. {\bf 94}, 102502 (2009).

\bibitem{deGraaf:PRL}
S. E. de Graaf, J. J. Leppakangas, A. Adamyan, A. V. Danilov, T. Lindstrom, M. Fogelstrom, T. Bauch, G. Johansson, and S. E. Kubatkin,
Phys. Rev. Lett. {\bf 111}, 137002 (2013).

\bibitem{reilly:natcomm}
J.~I. Colless, X.~G. Croot, T.~M. Stace, A.~C. Doherty, S.~D. Barrett, H.~Lu,
  A.~C. Gossard, and D.~J. Reilly, Nat. Comm. \textbf{5}, 3716 (2014).

\bibitem{Valenzuela08122006}
S.~O. Valenzuela, W.~D. Oliver, D.~M. Berns, K.~K. Berggren, L.~S. Levitov, and
  T.~P. Orlando, Science \textbf{314}, 1589 (2006).

\bibitem{PhysRevB.82.134522}
A.~Ferr\'{o}n, D.~Dom\'{i}nguez, and M.~J. S\'{a}nchez, Phys. Rev. B
  \textbf{82}, 134522 (2010).

\bibitem{PhysRevB.82.144524}
L.~Du and Y.~Yu, Phys. Rev. B \textbf{82}, 144524 (2010).

\bibitem{Laird2009}
E.~A. Laird, C.~Barthel, E.~I. Rashba, C.~M. Marcus, M.~P. Hanson, and A.~C.
  Gossard, Semicond. Sci. Technol. \textbf{24}, 064004 (2009).

\bibitem{PhysRevLett.112.227601}
J.~Stehlik, M.~D. Schroer, M.~Z. Maialle, M.~H. Degani, and J.~R. Petta,
Phys. Rev. Lett. \textbf{112}, 227601 (2014).

\bibitem{PhysRevB.84.241305}
E.~I. Rashba, Phys. Rev. B \textbf{84}, 241305 (2011).

\bibitem{PhysRevB.89.115409}
G.~Sz\'echenyi and A.~P\'alyi, Phys. Rev. B \textbf{89}, 115409 (2014).

\bibitem{PhysRevB.86.125428}
M.~P. Nowak, B.~Szafran, and F.~M. Peeters, Phys. Rev. B \textbf{86}, 125428
  (2012).

\bibitem{stehlik:v1}
J.~Stehlik, M.~D. Schroer, M.~Z. Maialle, M.~H. Degani, and J.~R. Petta,
  arXiv:1312.3875v1 (2013).

\bibitem{katja:science}
K.~C. Nowack, F.~H.~L. Koppens, Y.~V. Nazarov, and L.~M.~K. Vandersypen,
  Science \textbf{318}, 1430 (2007).

\bibitem{laird}
E.~A. Laird, C.~Barthel, E.~I. Rashba, C.~M. Marcus, M.~P. Hanson, and A.~C.
  Gossard, Phys. Rev. Lett. \textbf{99}, 246601 (2007).

\bibitem{footnote_fourthlevel}
Including a fourth level representing the second unpolarized (1,1) level would produce to first approximation another, equivalent series of resonances at $E' = n\omega$, where $E'/E = g_1/g_2$ is set by the ratio of the effective g-factors in the two dots. Adding the fifth level, $T_-$, would not significantly change the physics.

\bibitem{PhysRevB.81.201305}
S.~Nadj-Perge, S.~M. Frolov, J.~W.~W. van Tilburg, J.~Danon, Y.~V. Nazarov,
  R.~Algra, E.~P. A.~M. Bakkers, and L.~P. Kouwenhoven, Phys. Rev. B
  \textbf{81}, 201305 (2010).

\end{thebibliography}
\end{document}